\documentclass[ 4apaper,amsmath,amssymb ] {aipproc}
\layoutstyle{6x9}
\usepackage{graphicx}
\usepackage{bm}

\begin{document}

\title{Predicting and analyzing topological correlations in the CMB}
\author{A. Niarchou}{address={Astrophysics Group, Blackett Laboratory, Imperial College London, Prince Consort Road, London SW7 2AZ, UK}}
\author{A. H. Jaffe}{address={Astrophysics Group, Blackett Laboratory, Imperial College London, Prince Consort Road, London SW7 2AZ, UK}}

\date{today}

\begin{abstract}
The Cosmic Microwave Background (CMB) power spectrum derived from the first year WMAP data demonstrates an intriguing lack of power at large scales 
that cannot be accounted for within the framework of the standard cosmological model. We explore the possibility that this anomaly could be explained by 
waiving the implicit assumption of a simply connected topology, rather than modifying the physics of the standard model. In particular, we assume that the
Universe is slightly closed and its spatial section can be described by one of the simplest spherical multi-connected manifolds (the quaternionic, the 
octahedral, the truncated cube and the Poincare space). We discuss the implications for the CMB in each case. 
\end{abstract}

\maketitle

\section{Introduction}
\begin{figure}
\label{fig:discrepancy}
\resizebox{1.0\columnwidth}{!}
{\includegraphics{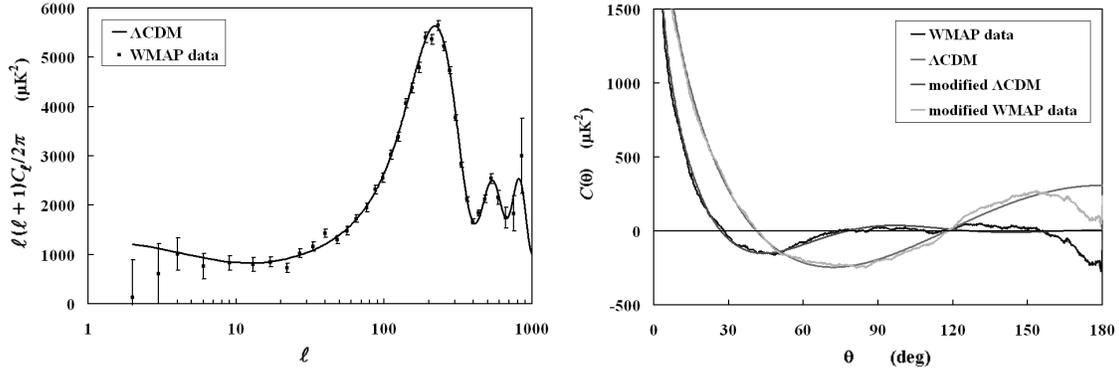}}
\caption{{\bf{Left:}}The CMB power spectrum for the WMAP best-fit model. There is a deficit of power at large scales (specifically ${\ell}=2, 3$). {\bf{Right:}} The
temperature anisotropy correlation function from the WMAP data, as predicted from the WMAP best-fit $\Lambda$CDM, for the best-fit $\Lambda$CDM with
$C_2$ and $C_3$ changed to the observed values and from the WMAP data with $C_2$ and $C_3$ matching the predictions of the best-fit $\Lambda$CDM. It
is clear that the disagreement between theory and observation is due to the lack of power at ${\ell}=2, 3$.}
\end{figure}

The CMB has proved to be an invaluable tool in our quest to unveil the mysteries of the cosmos. It dates back to the epoch of matter-radiation decoupling and the
expansion of the Universe has caused its mean temperature to drop to a mere 2.726 K today. It exhibits minute temperature fluctuations whose properties
depend on fundamental cosmological parameters and initial conditions. It is customary to expand the temperature fluctuation in a given direction $\hat{n}$ on the sky in a series
of spherical harmonics:
\begin{equation}
\label{eq:deltat}
\frac{{\Delta}T}{T}(\hat{n})=\sum_{\ell=0}^{\infty}\sum_{m=-\ell}^{\ell}a_{{\ell}m}Y_{{\ell}m}(\hat{n})
\end{equation}
and use the coefficients $a_{{\ell}m}$ to infer the parameters of interest. To this end, we usually employ their correlation, 
${\langle}a_{{\ell}m}a^*_{{\ell}'m'}\rangle$. In the case of a simply connected universe all off-diagonal terms are zero:
\begin{equation}
\label{eq:powersp}
{\langle}a_{{\ell}m}a^*_{{\ell}'m'}\rangle=C_{\ell}\delta_{\ell\ell'}\delta_{mm'}
\end{equation}
and we define the diagonal (rotationally invariant) part of the correlation matrix, $C_{\ell}$, to be the temperature anisotropy power spectrum. Another derived
quantity often used is the two-point correlation function, which can either be calculated directly from the CMB map, or constructed from the power spectrum:
\begin{equation}
\label{eq:correlation}
C(\theta)\equiv\left\langle\frac{{\Delta}T}{T}(\hat{m})\frac{{\Delta}T}{T}(\hat{n})\right\rangle_{\hat{m}\cdot\hat{n}=\cos{\theta}}=\frac{1}{4\pi}\sum_{\ell}C_{\ell}P_{\ell}(\cos{\theta})
\end{equation}
where $P_{\ell}$ are the Legendre polynomials. Since the aforementioned anisotropies are very small (of the order $10^{-5}$), extremely
accurate measurements of the CMB temperature are of the essence. The WMAP satellite, launched in 2001, has been the most accurate CMB experiment 
to date in the range of large to medium scales. Its first year data seem to be overall consistent with the standard cosmological model ($\Lambda$CDM), a
flat cold dark matter universe, dominated by dark energy and whose primordial perturbations are adiabatic and nearly scale-invariant \cite{wmap1}. However, there is a
striking disagreement between the predictions of that model and the actual data at large scales, which is prominent in the ${\ell}{\leq}3$ power spectrum and the correlation
function (Fig.~\ref{fig:discrepancy}).
There have been some attempts to remove this discrepancy by invoking new physics, but most of them result in contrived models with low statistical 
significance \cite{mypaper}. However, this phenomenon arises naturally in the context of non-trivial topologies, a possibility we will consider next.

\section{Spherical multi-connected manifolds}
\begin{figure}
\label{fig:platonic}
\resizebox{0.6\columnwidth}{!}
{\includegraphics{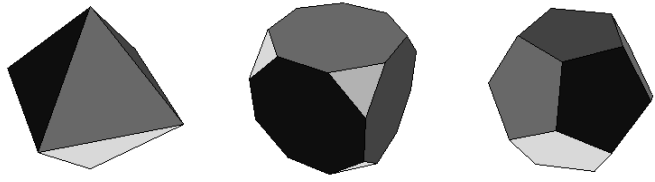}}
\caption{Fundamental domains of some spherical multi-connected manifolds. From left to right depicted are the regular octahedron, the truncated cube and
the regular dodecahedron.}
\end{figure}

A slightly closed universe is marginally consistent with the WMAP data (${\Omega}_0=1.02{\pm}0.02$). If it has a non-trivial topology, its spatial section could be
described by one of the spherical multi-connected manifolds.
Closed multi-connected manifolds are the quotient spaces ${\mathbf{S}}^3/\Gamma$, where ${\mathbf{S}}^3$ is the 3-sphere and $\Gamma$ is the symmetry
group particular to a given manifold. According to the way it acts on ${\mathbf{S}}^3$, we
obtain different categories of manifolds. We are only interested in some of the simplest manifolds (single action manifolds) \cite{spaces}:
\begin{itemize}
\item {\bf{the quaternionic space}}, where $\Gamma=D^*_2$, the binary dihedral group of order 4. The fundamental domain is a 4-sided prism.
\item {\bf{the octahedral space}}, where $\Gamma=T^*$, the binary tetrahedral group of order 24. The fundamental domain is a regular octahedron.
\item {\bf{the truncated cube space}}, where $\Gamma=O^*$, the binary octahedral group of order 48. The fundamental domain is a truncated cube.
\item {\bf{the Poincar\'{e} space}}, where $\Gamma=I^*$, the binary icosahedral group of order 120. The fundamental domain is a regular dodecahedron.
\end{itemize}
In each case, the number of fundamental domains that tile the 3-sphere is equal to the order of the group $\Gamma$. Figure \ref{fig:platonic} shows 
fundamental domains of some spherical multi-connected manifolds.

\section{Imprints of a non-trivial topology}
The most obvious effect of a non-trivial topology would be a suppression of power at large scales, due to the finiteness of the fundamental domain: evidently,
no wavelengths larger than the fundamental domain can exist, therefore the primordial power spectrum should be truncated at scales close to the size of 
the fundamental domain. The latter depends on the curvature radius, since the number of fundamental domains that tile the 3-sphere is constrained. Another
crucial effect would be the breakdown of global isotropy, as all the manifolds we consider are orientable. This would manifest as the emergence of non-diagonal
terms in the temperature anisotropy correlation matrix, appearing as non-Gaussianity. Table \ref{tab1} summarizes the differences between simply connected
and multi-connected spaces. We observe that the causal micro-physics governing the evolution of CMB anisotropies remain unchanged; what changes are the eigenmodes
of the Laplacian. This affects CMB observables in two ways: firstly, in multi-connected spaces not all wavenumbers ${\beta}{\geq}3$ correspond to Laplacian eigenmodes 
(in contrast to simply connected spaces) so that the ${\beta}$-contribution to each multipole is not the same; secondly,
since the eigenmodes affect the evolution of fluctuations along the line of sight, we expect the latter to evolve differently in spaces of different topology. 
The Laplacian eigenmodes of a multi-connected space are the $\Gamma$-invariant eigenmodes of its universal covering space and they
can be expressed as a series of the actual eigenmodes of the latter. Thus, in spherical multi-connected manifolds, they can be expressed on a basis of the
eigenmodes of ${\mathbf{S}}^3$ (e.g. as a product of a hyperspherical Bessel function and a spherical harmonic). The expansion coefficients ${\xi}_{\beta{\ell}m}$
are particular to a given topology. The effect of that topology on the CMB is expressed through the presence of the ${\xi}_{\beta{\ell}m}$ in the correlation
matrix.
  
\begin{table}
\begin{tabular}{cc}
\hline
\tablehead{1}{c}{c}{simply connected}
&\tablehead{1}{c}{c}{multi-connected}\\
\hline\hline
\rule[-3mm]{0mm}{8mm}${\cal{Y}}_{\beta}({\chi},{\theta},{\phi})$&${\Psi}^s_{\beta}({\chi},{\theta},{\phi})=\sum_{{\ell}=0}^{{\beta}-1}\sum_{m=-\ell}^{\ell}{\xi}_{{\beta}{\ell}m}^s{\cal{Y}}_{\beta}({\chi},{\theta},{\phi})$\\
\hline
\rule[-3mm]{0mm}{8mm}$a_{{\ell}m}=i^{\ell}{\int}{\beta}^2\sqrt{P(\beta)}{\Delta}_{T\ell}(\beta)e_{\beta{\ell}m}^sd{\beta}$&$a_{{\ell}m}=\frac{(2\pi)^3}{V}\sum_{\beta}\sqrt{P(\beta)}{\Delta}_{T\ell}(\beta){\sum}_s{\xi}_{\beta{\ell}m}^se_{{\beta}s}$\\
\hline
\rule[-3mm]{0mm}{8mm}${\langle}a_{{\ell}m}a_{{\ell}'m'}^*{\rangle}=C_{\ell}{\delta}_{\ell{\ell}'}{\delta}_{mm'}$&${\langle}a_{{\ell}m}a_{{\ell}'m'}^*{\rangle}=C_{{\ell}m}^{{\ell}'m'}\;\;\;\;C_{\ell}=\frac{1}{2\ell+1}\sum_{m=-\ell}^{\ell}C_{{\ell}m}^{{\ell}m}$\\
\hline
\end{tabular}
\caption{Effects of a multi-connected topology. The eigenmodes of the Laplacian can be decomposed on a basis of the eigenmodes in the universal covering
space, with the expansion coefficients encoding the topological properties of the given space (first row) \cite{ghost,multimaps}. We see that the correlation matrix is no longer
diagonal in the multi-connected case, however a rotationally invariant power spectrum can still be defined \cite{multimaps}. Note that the physics remain unchanged and all
effects of a non-trivial topology are expressed through the coefficients ${\xi}_{\beta{\ell}m}$. Here $\cal{Y}$ and $\Psi$ are the eigenmodes of the 
universal covering space and the multi-connected space respectively, $P(\beta)$ is the primordial power spectrum and ${\Delta}_{T\ell}$ are the transfer functions.
s is the multiplicity of a given eigenmode, V is the volume of the fundamental domain and the e's are uncorrelated, appropriately normalized Gaussian random
variables.}
\label{tab1}
\end{table}

\begin{figure}[h]
\label{fig:maps}
\resizebox{0.9\columnwidth}{!}
{\includegraphics{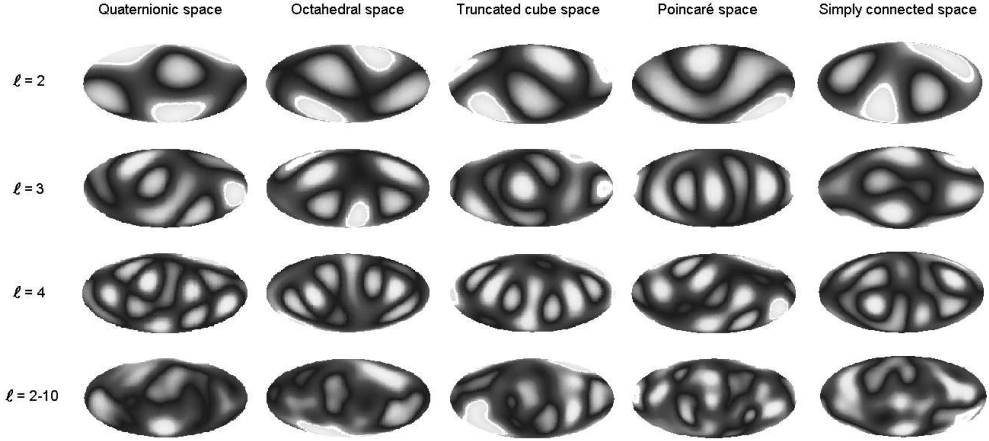}}
\caption{CMB sky realizations for low multipoles for a closed $\Lambda$CDM universe with simply/multi-connected topology.}
\end{figure}

\begin{figure}
\label{fig:powersp}
\resizebox{0.75\columnwidth}{!}
{\includegraphics{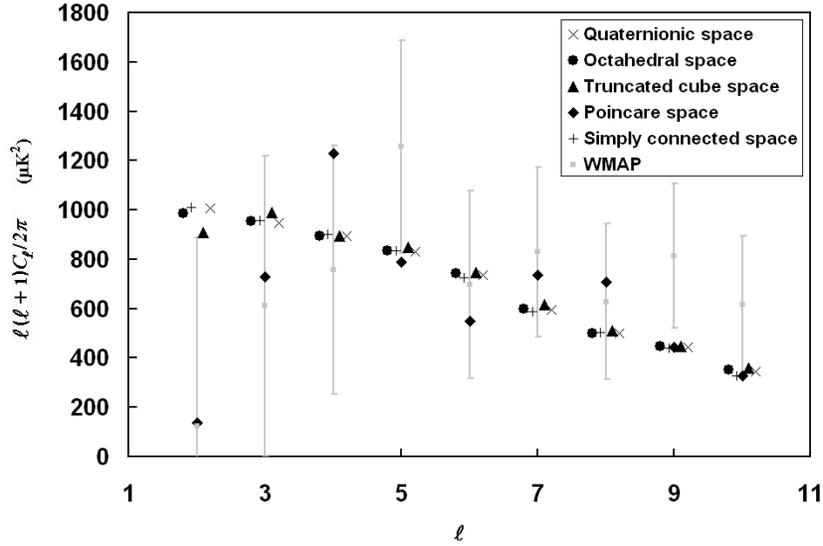}}
\caption{CMB power spectra for closed $\Lambda$CDM universes. The smaller the fundamental domain the larger the suppression at low $\ell$: the power spectra of the simply connected, the 
quaternionic and the octahedral spaces almost coincide, while that of the truncated cube space has visibly less power. The power deficit is most prominent in the Poincar\'{e} space. Note that
some points have been slightly displaced from their original x-position for reasons of clarity.}
\end{figure}

Figures \ref{fig:maps} and \ref{fig:powersp} show CMB sky realizations and power spectra for a closed universe with simply/multi-connected topology and
with ${\Omega}_b=0.057$, ${\Omega}_{CDM}=0.275$, ${\Omega}_{\Lambda}=0.683$, ${\Omega}_k=-0.015$
and $H_0$=65 km/sec/Mpc. Note that the drop at higher $\ell$ in Fig.~\ref{fig:powersp} is partly artificial and due to the fact that only wavenumbers ${\beta}<33$(${\sim}\,0.88$/kpc) were used in
the calculation. The simply connected universe has the same power spectrum as the WMAP best-fit at large scales and since it matches those of the multi-connected spaces from ${\ell}\,{\simeq}\,9$ onwards, we expect the latter to coincide with the power spectrum of the WMAP best-fit at smaller scales. Finally, we observe that spaces with smaller fundamental domains exhibit greater suppression of power at low multipoles, while non-trivial topologies can also induce features such as moment alignment in CMB maps.

\section{Future prospects}
Our calculations show that spherical non-trivial topologies reproduce the observed low power at large scales by construction. In addition, they can lead to
moment alignment, a feature claimed to be present in the WMAP data. Therefore, multi-connected topologies seem a viable scenario that could explain the intriguing findings 
of WMAP, especially since there are no known constraints on the topology of the Universe. 
The intrinsic anisotropy of the manifolds in question forces us to consider the full correlation matrix and not just its diagonal
part, enabling us to gain better insight into the physics of large scales. We are now calculating the likelihood functions for these topological spaces, in order
to assess their statistical significance by means of Bayesian model comparison.

\newcommand{\apjs}{ApJS}
\newcommand{\prd}{Phys.Rev.D}
\bibliographystyle{aipproc}
\bibliography{elaset.bib}
\end{document}